# Who Invented the Trinity Nuclear Test's Christy Gadget?
# Patents and Evidence from the Archives

Thomas A. Chadwick* and M.B. Chadwick
Los Alamos National Laboratory
Los Alamos, NM 87545

**Abstract**: The Christy Gadget is the informal name for the plutonium device detonated in the Trinity test on July 16, 1945. In September 1944, Robert Christy, working in the theoretical implosion group, proposed a novel concept that altered the design of the nuclear core in Fat Man. While scientists originally intended to use a hollow sphere of plutonium, this design entailed substantial risk, due to the likelihood of asymmetries resulting from implosion. Christy proposed changing the design to a solid sphere of plutonium with a modulated neutron source, and the design was eventually adopted, tested at Trinity, and used in the attack on Nagasaki. While there is no question regarding the important role that Christy played in demonstrating its feasibility as a reliable design, there is a debate as to who initially proposed the idea; though most sources have attributed this invention to Christy, some historical sources have attributed credit to Christy's group leader, Rudolf Peierls, or indeed other scientists. This paper seeks to outline and resolve this dispute. We present new unclassified evidence extracted from previously unavailable sources (to unclassified audiences) from the National Security Research Center archives at Los Alamos National Laboratory. This evidence consists of 1945–1946 patent documentation, oral history interview tapes of Christy and Peierls, and monthly 1944 progress reports from the Theoretical Division. Though Christy and Peierls share joint credit on the patent, both Christy's and Peierls' words and writings, together with sources from Hans Bethe and Edward Teller, support the traditional view that Christy was indeed the originator of the idea. While Christy does deserve the majority of the credit for the invention and design, we acknowledge the important role Peierls and von Neumann played in its development.

## 1. Introduction

The Trinity test, performed on July 16, 1945, was the first detonation of an atomic weapon. The test was a remarkable achievement: it was an amalgamation of numerous scientific discoveries and inventions and it brought the war-torn world into a new atomic age. Fundamentally, it proved the feasibility of a plutonium implosion design that would be subsequently used in the Fat Man device detonated over Nagasaki. Robert Christy, a scientist working in the Theoretical Division during the Manhattan Project, has his name firmly tied to the Trinity test. The gadget tested in Trinity is informally known as the "Christy Gadget" because Christy made an important, novel proposal in September 1944 for a new design of the plutonium core.

Until late 1944, scientists at the Manhattan Project hoped to be able to use a hollow plutonium shell as the core of the implosion device. Upon detonation, the high explosives would create an imploding shock wave that would compress the hollow plutonium shell into a super-critical mass, creating a nuclear yield. While this design was elegant and efficient, in practice, after extensive experimentation it was still plagued by asymmetries upon implosion, including jetting, spalling, and Rayleigh-Taylor instabilities. Scientists feared that these asymmetries might prevent the reliable assembly of a critical mass.

Consequently, many scientists remained skeptical about the feasibility of successfully building an implosion device on the urgent timeframe needed to impact the war effort. Segrè's discovery that reactor-produced plutonium had too-high a spontaneous neutron emission, preventing a gun-type Thin Man assembly design, and the challenges of producing large quantities of enriched uranium, elevated the implosion design to the center of the laboratory mission. Oppenheimer reorganized the laboratory in August 1944 to orient the laboratory mission around the implosion device, establishing two divisions, the Weapons Physics or Gadget Division (G) and the Explosive Division (X), both intended to primarily focus on the implosion design.[1,2] As a sign of the urgency, when Arthur Compton visited the laboratory on August 1, 1944, he gave a colloquium and advised the audience that they had already used "*half of the time of the estimated maximum to produce a successful gadget*."[3]

* thomaschadwick@berkeley.edu

In September 1944, Christy proposed a new conservative design for the plutonium pit, suggesting a change from the hollow plutonium shell to a solid sphere of plutonium with a modulated neutron source. The solid sphere of plutonium was surrounded by a tamper and explosives that would simultaneously explode to form a convergent detonation wave, compressing the plutonium symmetrically. The modulated neutron source refers to the concept that a source was desired that provided very few neutrons at the beginning of the implosion (to avoid pre-initiation) but a large number of neutrons later when the plutonium would be compressed. This design relied on assembling a near-critical mass of plutonium before compressing it, increasing its density and creating a supercritical assembly. While Christy's new invention was thought to be less efficient than the hollow shell, the solid sphere was a more reliable design because it minimized the implosion asymmetries. Although there is no question that Christy performed important hydrodynamic and nuclear research to prove the feasibility of this design, there is a debate in the open literature as to who invented the concept. As will be discussed further, the 1946 patent is in the names of both Christy and Peierls. While a number of sources make claims about the origins of this idea, we have yet to find any source which summarizes the evidence into an argument about the genesis of the Christy Gadget. The purpose of this paper is to review evidence from both open and classified sources, and to provide unclassified extracts from the National Security Research Center (NSRC) at Los Alamos to answer this question.

While Christy's name has become synonymous with this novel design, many different sources attribute credit for this invention to scientists other than Robert Christy. In *Britain and the H-Bomb*, British historian Lorna Arnold claims that the solid sphere implosion design was actually invented by Rudolf Peierls, the Hydrodynamics of Implosion group leader and a member of the British Mission to Los Alamos.[4] Arnold's claim is referenced by Oxford physicist, Frank Close, in his excellent 2019 book *Trinity: The Treachery and Pursuit of the Most Dangerous Spy in History*. According to Close, an "authoritative history" credits Peierls with the invention.[5] Even figures like Edward Teller, one of the key inventors of the hydrogen bomb, and John von Neumann, the Hungarian mathematician who made numerous innovations on implosion and high-explosive lens systems, have been credited with inventing the solid-sphere implosion design.

While some sources attribute credit for the invention to scientists other than Christy, the evidence for these claims is weak. As we shall show, the references given for Arnold's claim do not support the view that Peierls was the original inventor. Furthermore, the idea that Teller or von Neumann was the principal originator of the Christy Gadget design is not supported by evidence. Indeed, none of these remarkable scientists actually made this claim themselves. While we recognize that this invention required a collaborative effort, we conclude that the name of this invention is aptly bestowed and argue that Robert Christy deserves principal credit for this invention.

## 2. Dramatis Personae

### *A. Robert Christy*

Robert Christy (Fig. 1), a Canadian who was naturalized as a US citizen, did his PhD work under J. Robert Oppenheimer at the University of California, Berkeley, before moving to the University of Chicago to work on the reactor experiments with Fermi and Wigner. He was an early recruit to the Manhattan Project, arriving in Los Alamos in 1943.

Christy helped work on the water boiler reactor and then worked in the Theoretical Division's Hydrodynamics of Implosion group (T-1), first under Edward Teller and then under Rudolf Peierls. In June 1944, Oppenheimer and the Theoretical Division Leader Hans Bethe transferred Teller from the T-1 leadership role to instead focus on thermonuclear concepts (the "Super"). In the 1986 oral history interview with Hoddeson[6] (see Sec. 4), Christy couldn't remember that Teller was his group leader, stating, *"I don't remember working under Teller"* and *"Mostly he pursued his own thing. He wasn't much of a team player"*![†] Teller himself noted that he wasn't the right person to lead the calculational implosion effort, and that Bethe's initial desire for Teller to lead this needed a different solution. When the British scientists started arriving in Los Alamos, Teller said, "*Bethe saw Peierls as ready-made for the task he had in mind for me. Peierls could tackle the calculations of implosion. So Peierls, with the help of a small group, diligently began the Herculean labor.*"[7]

After Los Alamos, Christy had a scientific research and teaching career at Caltech and served as President of Caltech for a period.

### *B. Rudolf Peierls*

Rudolf Peierls (Fig. 2) was a German-born physicist known for his expertise in nuclear physics, material science, and shock hydrodynamics. In Germany, he studied physics under Heisenberg and Pauli before moving to the Cavendish Laboratory in Cambridge. He became a naturalized British citizen and after the war was knighted. Peierls was a coauthor of the famous March 1940 Frisch-Peierls memorandum. This in turn led to the creation of the MAUD committee and its report,[8] drafted by Sir James Chadwick, which expedited research in atomic weapons in Britain and in the USA in the early 1940s. Peierls came to Los Alamos in 1944 as part of the British Mission, becoming leader of the Theoretical Division's T-1 group.

[†] Following the Oppenheimer security hearing in the 1950s, Christy had cold relations with Teller. Teller's memoirs describe Christy spurning him (Ref. 7).

It is perhaps useful to provide some background that illuminates Peierls' brilliance and the extent of his essential contribution to the Manhattan project.

In 1942, Oppenheimer was already communicating with Peierls by letter on detailed questions in nuclear science and weapons physics.[9] As part of his goal to recruit Peierls to Los Alamos, Oppenheimer wrote a letter to Leslie Groves on February 14, 1944, in which he praised Peierls' expertise in the hydrodynamics of implosion.[10] Oppenheimer noted that the British had some complementary technical knowledge of blast waves that was needed in Los Alamos[11].

On July 18, 1949, Bethe wrote a letter to Carroll L. Wilson, general manager of the Atomic Energy Commission, documenting the role of various British scientists. Of Peierls, he said, "*He joined the Los Alamos Project at a time when it was impossible to find a competent senior theoretical physicist to head the group without crippling some other phases of the work. Peierls was, in my opinion, the most effective group leader in the Theoretical Division. . . . He directed the entire difficult work on implosion hydrodynamics, contributing a very great fraction of it himself. . . . Due to the work of the group the project had sufficient theoretical knowledge of the implosion to feel confident of the feasibility of the implosion weapon*." Director Norris Bradbury, in a letter on the same date also to Wilson, stated, "*It should be noted that the British Mission supplied the major portion of experience in the field of theoretical hydro-dynamics . . . the U.S. was largely lacking in personnel experienced in this field of classical physics.*"[12]

Frank Close's book refers to Peierls as the father of the atomic bomb, a title that is traditionally also applied to J. Robert Oppenheimer. This obtains from his joint authorship on the Christy gadget patent and his shock hydrodynamic calculation leadership at Los Alamos. It also reflects his early 1939 Cambridge Philosophical Society paper that derives a critical mass formula for unmoderated fast-neutron systems and the Frisch-Peierls 1940 memorandum that first computed the critical mass and efficiency of an enriched uranium bomb.[13]

### *C. John von Neumann*

An émigré from Hungary to Princeton, the brilliant mathematician and theoretical physicist came to Los Alamos from late 1943 onwards as a consultant, at Oppenheimer's invitation.

Von Neumann (Fig. 3) was a renowned expert in shock hydrodynamics and pioneered the development of computing on electronic machines. Stan Ulam's obituary of von Neumann[14] spoke of his "*ability, perhaps some-what rare among mathematicians, to commune with the physicists, understand their language, and to transform it almost instantly into a mathematician's schemes and expressions. Then, after following the problems as such, he could translate them back into expressions in common use among physicists.*" Von Neumann suggested using large amounts of high explosives to cause exceedingly high velocities in an implosion, and both von Neumann and Teller had the insight that this would increase compression to create a critical mass (Ref. 7, p. 175). Together with James Tuck (who had been working on related shaped-charge research in England[15]) and Seth Neddermeyer, he suggested that a symmetric implosion could be accomplished using a high-explosive lens system.

### *D. Ralph Carlisle Smith*

Major Ralph Carlisle Smith (Fig. 4) came to Los Alamos in 1944. He was an Office of Scientific Research and Development (OSRD) representative tasked by Oppenheimer to manage patents during the Manhattan Project. He subsequently became an assistant director of the Laboratory from 1947 to 1957. He later moved to become the president of New Mexico Highlands University. He was a patent attorney with a first degree in chemical engineering from the Rensselaer Polytechnic Institute.

An organizational chart of the Theoretical Division, May 10, 1945, is shown in Fig. 5, showing Peierls and Christy in T-1 as well as listing numerous other luminaries. (Note that the Los Alamos WRL journal's 2021 special issue on Trinity has a paper by Shlachter describing the contributions of the Jewish members of T-Division.)

## 3. Discrepancies Over the Inventor of the Christy Gadget

Lorna Arnold, in her book *Britain and the H-Bomb*, asserts that Peierls came up with the idea of the Christy Gadget (Ref. 4, p. 254). While Arnold's book mainly examines the British efforts to build a hydrogen bomb at Aldermaston in the 1950s, she also discusses the contributions of the British Mission at Los Alamos. In a footnote, Arnold claims that the "Christy Gadget" was actually Peierls' idea (Ref. 4, p. 254). It is worth quoting Arnold in full:

*"At Los Alamos, Peierls and Fuchs provided two-thirds of the team which made the implosion development possible and contributed to all phases of weapons development (including the Super). The solid implosion gadget invented by Peierls and Christy is commonly called the Christy gadget* ***but was Peierls' idea****. Tuck, independently and with the US scientists Neddermeyer and von Neumann, suggested the lens system for implosion and worked with Bethe on the initiator. Frisch made many contributions, especially to critical mass assembly studies. Bretscher made considerable contributions to Super feasibility studies. Titterton did outstanding work, particularly on electronic circuit developments. Rotblat worked with several others in the field of experimental nuclear physics. See F. Szasz, British Scientists and the Manhattan Project, pp. 148–51."*

It should be noted that Arnold's citation is to a book entitled *British Scientists and the Manhattan Project* by Szasz. However, the information in Szasz's book is actually the appendix that reproduces a memorandum by Ralph Carlisle Smith, a memorandum that is discussed below. It represents a fair reproduction, except for the words "*but was Peierls' idea.*"‡ Close's new book *Trinity: The Treachery and Pursuit of the Most Dangerous Spy in History* cites Arnold's claim (Ref. 5 p. 130 and his Reference 13, p. 451).

Arnold's claim contradicts other sources, most notably *Critical Assembly*, a book by several Los Alamos historians,[16] and deserve some attention. She cites only one reference to support the claim, the 1949 memorandum written by Ralph Carlisle Smith.[17] The evidence from LANL's patent collection indicates that Smith was intimately involved with the patent creation process, as many of the drafts and administrative papers bear his signature. At the conclusion of the project, Smith wrote a summary of the contributions of the British Mission, with a concise description of each of their contributions to the project. It is clear that he takes great care to communicate fairly the credit due to the various scientists.

But Smith's summary in no way corroborates Arnold's claim. Smith writes that Peierls and Christy were "joint inventors" of the Christy Gadget, providing both with credit for the invention (as is appropriate since both are named on the patent, see Sec. 4A). Smith also notes that Peierls was an expert in hydrodynamics and that his work was crucial for the development of the implosion device, contributing to "*all phases of weapon development, including implosion and Super*."[18] While this document does give Peierls a role in the invention of the Christy Gadget and commends his expertise in hydrodynamics, no statement in Smith's summary corroborates the claim that Peierls originated the idea.§

### *A. Invention Claims Attributed to John von Neumann and Edward Teller*

Smith's memorandum on the contributions of the British Mission also gives some credit to John von Neumann for first developing the idea of the solid-sphere design. He ascribes the "Christy Gadget" invention jointly to Peierls and Christy, but also states that "*Solid implosion had been suggested by von Neumann in his early patent application but the idea of a modulated neutron source with the solid implosion was that of Peierls and Christy.*"

It is true that the Christy-Peierls Patent S-3956X gives "prior art" as von Neumann's 1943 patent S-673X (Fig. 6). The key contribution of von Neumann's patent was to propose using a large mass of high explosive to cause plutonium compression. Von Neumann discusses numerous concepts that include a hollow shell of plutonium (later called the "standard gadget") as well as designs with a solid spherical plutonium component. However, his patent does not include the eventual solid-sphere Christy gadget design per se. When von Neumann discussed a solid component, the design configuration differed from the Christy Gadget design, in that it included an air gap and more complex components in addition to its "levitated" solid sphere. While these designs were used in later Los Alamos tests after the war, they lacked the simplicity and robustness of the Christy-Peierls invention. Von Neumann's design was one of the many concepts proposed during the Manhattan Project, but it is inaccurate to label these design ideas as synonymous with the Christy Gadget.

Thus, our assessment from reading von Neumann's patent is that Ralph Carlisle Smith's explanation was not quite right. He slightly mischaracterized the essential novel idea of Peierls and Christy as a *modulated source* (the neutron initiator) with solid sphere, instead of also emphasizing the lack of free space within the explosives.

It has also been suggested that Edward Teller invented the idea of a solid-sphere implosion device. Alex

‡ Arnold's statement of the British contribution to implosion being "*two-thirds*" (the third being von Neumann or Christy, presumably) is from Smith's memorandum.

§ We think that Smith wrote "Peierls" before "Christy" simply because his entire paragraph was focused on Peierls' role in the Manhattan Project, and not because he thought Peierls originated the idea.

Wellerstein,[19] a history of science professor at the Stevens Institute of Technology, states:

"*The solid-core concept was originally proposed by Edward Teller, whose experience working with George Gamow on the iron core of the Earth gave him the somewhat unintuitive knowledge that even very dense materials can be compressed to even higher densities under many megabars of uniform force. Christy, for his part, was the guy who took it from the 'Teller's interesting but potentially wrong idea' phase to the 'so will it actually work?' phase. And he did a good job of that—everyone was convinced that a solid core bomb would be both plausible and easier than the alternatives (such as a hollow core bomb).*"

However, although Teller's *Memoirs* (Ref. 7, p. 175, see Sec. 5B) does describe the increased compression that occurs under pressure and these memoirs provide the analogy of the Earth's iron core, this is given in the context of the original "standard gadget" shell implosion and not in the solid ball design. Our reports in the NSRC do show a Teller patent (S1201X, Nov. 1, 1945, Box 3) that includes a levitated solid sphere, but like von Neumann's 1943 patent, it also includes an air gap. In fact, Teller's own words, reproduced below in Sec. 5B, clearly show that he credits Christy with the "ingenious" solid ball invention.

## 4. Sources from Los Alamos NSRC

### *A. Patent Collections*

During the Manhattan Project, scientists were encouraged to document their innovations in patent applications, with the assistance of Ralph Carlisle Smith. Professor Sabine Lee noted that "*Political leaders . . . had been discussing the question of intellectual ownership and commercial rights throughout the war, and they considered the implications of individual countries' contributions to the international nuclear venture as significant in view of post-war bargaining.*"[20]

These patents (final versions and early drafts), together with extensive related documentation and correspondence, exist in the NSRC. There are numerous patents related to implosion concepts in the names of well-known scientists such as von Neumann, Bethe, Neddermeyer, Teller, Fuchs, as well as Christy and Peierls, dating from 1943 on. Robert Christy and Rudolf Peierls jointly filed a patent in January 1946 entitled "Method and Apparatus for Explosively Releasing Nuclear Energy," which patented the idea of the "Christy Gadget." While some parts of the patent documentation still cannot be released, we have been able to make some part of the patent openly available here, and it remains one of the most exciting pieces of new evidence on this topic. In particular, Fig. 7 provides one example of the claim that appears in the patent. The paragraph begins with "It is claimed", and then proceed to lay out the characteristics of the solid-sphere implosion device.

The patent describes the implosion design of a solid plutonium sphere (without an air gap) with a modulated neutron source. The modulated neutron source was intended to selectively supply a large number of neutrons at the time of compression to initiate a nuclear chain reaction. Alternative external initiator concepts were discussed for shell implosions before the Christy Gadget proposal. The idea of a modulated neutron source at the center of the gadget was presented in the early 1945 Christy-only version of the patent, but R.C. Smith's previously quoted statement, "*but the idea of a modulated neutron source with the solid implosion was that of Peierls and Christy*," suggests that both Christy and Peierls both played a role in this aspect of the invention.

For each individual patent, the NSRC maintains handwritten and typed rough drafts, in addition to the final version of the patent, all of which provide a glimpse at the process behind creating the patent. These sets of documents provide strong evidence that Christy should be recognized as the originator of the invention. Christy wrote out the first draft in 1945, *with just his name on the document as sole inventor*. Figure 8, the earlier Christy-only 1945 version, shows the invention to have occurred "prior to October 2, 1944," and cites Christy's notebook A147 documenting the invention.

After the war, this version was given to Peierls for his review. Peierls made a number of small revisions by hand to the typed document and wrote "and Peierls" next to Christy's name (Fig. 9). The patent documents detail the transition from Christy's hand-written draft to the subsequent version with Peierls' name added. Then, the final document was prepared with both Christy's and Peierls' names, which appeared on the final patent submitted to R.C. Smith.[21]**

A number of possible conclusions can be drawn by re-creating the history of the patent. The most likely scenario is that Christy originated the idea and wrote the first draft, but sometime after this, discussions occurred between Christy, Peierls, and possibly Smith and others, leading to the view that the fairest course of action was to add

** The final Christy-Peierls 1946 patent lists Peierls as a subject of the King of Great Britain and they were paid one dollar for transferring their rights and title of this invention to the US Government.

Peierls' name to recognize his contributions. This hypothesis is expanded upon in our conclusions. There is also an indication in Smith's 1946 letter to Christy that discussions with government representatives in Washington also played a role in adding Peierls (Fig. 10).

Some additional images from the patent records are given in our report[22].

### *B. Oral History Tapes*

**Christy Interview**

In 1986, LANL historian Lillian Hoddeson conducted an interview with Robert Christy, in which he describes his role in the invention and development of the implosion device. He comes across as transparent, and not boastful, when he repeatedly uses the first person singular—"I" not "we"—in describing his insights and calculations.

In his remarks, Christy remembers putting forward the proposal for the Christy Gadget and calculated the efficiency of the solid-sphere design, finding it to be "reasonably efficient." Christy's proposal assuaged skeptical colleagues who doubted the feasibility of the implosion design.

He said, "*I do remember putting forward the proposal of an implosion of a solid. . . . I don't know where the idea came from, I just did it.*" Here, he is ruminating on the mysteries of how the creation idea originated in his own mind. (And in fact, as described in Sec. 4D, in the patent Christy *does* describe how this idea originated in his thinking.) Below, we quote verbatim unclassified text from that interview:

**Christy**: "*I was a believer in implosion simply because from the point of view of theory it was ideal, it was absolutely perfect. The only difficulty was, it wasn't sure you could make it go. But that didn't bother me as a theorist. From the theoretical point of view it had every advantage.*"

. . .

**Christy**: "*I don't remember when the name 'Christy Gadget' was first attached to it. I do remember putting forward the proposal of an implosion of a solid. . . . I don't know where the idea came from except that there were all these worries about what happens at the inside surface, spalling, jetting and so forth, and so at one point—I don't know exactly what preceded it—I tried to do it for a solid sphere, to see whether that would be a possibility, and it seemed as though it was. That is, it gave a reasonable efficiency. But I don't remember where the idea came from, I just did it. And, I was always a strong proponent of the implosion method and I guess I was always worried at all of the troubles that there were, and all of the asymmetries, and jetting, and spalling and that . . . . It worried me that people might discard this beautiful technique just because of the practical problems. So I was then pleased to devise a proposal that convinced even the skeptics that the implosion could work . . . and that's the way I viewed it, as a, as kind of the worst implosion you could design, one without any hole in the middle, but that would convince the doubters that they could go ahead with implosion.*"

**Hoddeson**: "*Do you remember the responses initially to your suggestion?*"

**Christy**: "*Well I remember talking to various people including Oppenheimer and I think the responses were favorable, namely that it looked as though you could hardly develop asymmetries in a solid sphere and the real question was whether you could—there was no place for them to go—and so the real question was whether or not the implosion gave you a sufficient compression so that you would get a real significant yield, and I had made calculations based on some of the approximate calculations I did that indicated it would, and then there were various measurements made by RaLa and X-Ray methods and so forth, and again, these also confirmed that there was a real compression, that looked as though it would work. So I think things fell in line pretty well because the various experimental tests agreed pretty well with the calculations and it looked as though it would really work.*"

**Peierls Interview**

Lillian Hoddeson also conducted an interview with Rudolf Peierls that same year.[23] When Hoddeson asks Peierls, "*Had Christy not come up with the solid sphere idea . . . do you think that there would have ultimately been a successful shell implosion device given what you know about the results?*" her phrasing of the question clearly ascribes credit to Christy for the invention of the Christy Gadget design. Peierls' response is telling: instead of contradicting her, he simply answers the question. Hoddeson then asks a question regarding the innovative nature of the Christy Gadget. Once again, Hoddeson's use of language is telling. She asks, "*Also, had Christy not come up with the idea of a solid . . . ?*" clearly attributing credit to Christy. Once again, Peierls answers the question without contradiction. Peierls' tacit acceptance of Hoddeson's questions provides evidence that he does not consider himself to be the principal inventor of the gadget, and towards the end, he states, "*Now, Christy's idea was to . . .*" Again, we quote verbatim unclassified extracts from that interview, where Peierls discusses Bethe's Theoretical Division progress reports from December 1944 and January 1945:

**Peierls**: "*This is the first time that numerical work on the IBM machines has been done for the Christy Gadget. So,*

*it took that long for this to be taken that seriously, because all the other numerical calculations still related to the hollow shell.*"

**Hoddeson**: "*Why do you think it took so long?*"

**Peierls**: "*Why it took so long to adapt the IBM calculations to the Christy gadget? Well, I think the answer must have been that it took a long time before the decision was reached that the Christy Gadget was the main thing to aim at.*"

**Hoddeson**: "*I don't think that decision was taken till February.*"

**Peierls**: "*Obviously before taking that decision, you would have wanted to see the results of the IBM calculations. But until then it was probably not a sufficiently serious candidate. And also, there was probably the point that they had started the calculations on the other gadget and obviously they should be finished before you changed horses.*"

**Hoddeson** (quoting Bethe's summary): "*Encouraged by the RaLa results, we have felt it desirable to make immediately an investigation of a solid-sphere implosion by an IBM computer.*"

**Peierls** (on his report from the end of December 1944 that became LAMS-182 on the problem of asymmetries in implosion, not for the solid sphere, but for the standard implosion): "*This is still relating to the hollow shell . . . Tuck's X-ray shots are mentioned, which are particularly good at showing up jets, and Koski's. But in general, it seems first of all that it is possible to avoid jets by using lenses, and once you have no jets the collapse of a spherical shell should not be too unstable. So, this paints a fairly optimistic picture, but with various reservations. . . . But still, leaving various problems to be solved practically, like getting lenses that are sufficiently well designed and sufficiently uniform and so on.*"

**Peierls**: "*This report comes to a very serious recommendation of the solid gadget, in January. . . . There is an interesting remark about why it was relatively late—the IBM calculations turned over to the Christy gadget . . . one doesn't appreciate today how slow these calculations were. . . . In January the IBM machine completed the 7$^{th}$ problem that they had done since they started, which was in April. The solid gadget was problem number 8. . . . Each calculation is almost a month. Bethe's progress report then mentions that the decision was frozen: the Christy gadget with initiator. The decision was taken during the month.*"

**Hoddeson**: "*Had Christy not come up with the solid sphere idea . . . do you think that there would have ultimately been a successful shell implosion device given what you know about the results?*"

**Peierls**: "*Well there were sufficient doubts. I mean, obviously one would have continued exploring all these questions of instabilities and spalling and jets and so on, and probably one could have reduced the uncertainties sufficiently, but one would be, in the final Trinity type test, one would have been less confident than with a solid gadget.*"

**Hoddeson**: "*Also, had Christy not come up with the idea of a solid . . . was it an obvious enough step then to take to go in that direction. . . . How much of an innovation was that suggestion? Would it probably have been something that eventually one of the other people working might have come to?*"

**Peierls**: "*It is very hard to say. I mean it did certainly come as a surprise. To say well, why do we have to have a hollow shell? Why can't we just . . . . You see, there is a gradual transition. The first purpose of the implosion idea of a hollow shell was simply a very fast way of assembling the material from the subcritical into a critical stage and there, the compression, the idea of compression, didn't play a large part. Then one noticed in the course of doing that, that one could gain considerably in compression, and therefore in density and therefore in criticality. But one thought of this increasing density as a result of all the material collapsing together, and therefore its kinetic energy being converted to potential energy. . . . So, you had three stages of the explosion of the detonation wave being converted to the kinetic energy of the shell and then as the collapse, the kinetic energy turning into potential energy and therefore compression. Now, Christy's idea was to notice that you could leave out of the intermediate stage and convert the energy of the detonation wave directly into potential energy. That was a very novel thought. Now, how likely it is that someone else would have come up with that, I just don't know.*"

Christy's invention is certainly novel, and worthy of a patent! The interview above provides clear evidence that Peierls viewed Christy as the principal originator of the solid-sphere invention.

### *C. Theoretical Division Monthly Progress Reports*

Perhaps the first reference to the solid Gadget concept, beyond Christy's notebook, was given in the Theoretical Division's September 1944 monthly progress report LAMS-149[24], see Fig. 11. Bethe introduces this new work

"*as a possible further insurance against failure to have symmetry in an implosion of more standard design.*" Subsequently, each group leader provided a summary of progress in their group. Peierls writes for T-1: "*Christy has recently made some estimates on the performance of a solid gadget."*(Fig. 11).

Subsequent 1944–1945 Los Alamos reports in the NSRC continue to refer to the Christy Gadget. The Gadget Physics (G) Division, led by Bacher, reported in its first progress report, Oct. 1, 1944, that "*an alternative form of limited objective is now being considered quantitatively by Christy, viz a gadget that is integrated in compact solid sphere form with tamper and then compressed by H.E.*" On Oct. 30, in LA-164, Christy and Bethe presented calculations of a solid-sphere implosion as a limiting case of shell implosions, noting that the "*compression will be finite even for a solid sphere.*" Experimental implosion work on compressions that could be achieved was reported in the Nov. 16, 1944, G Division progress report, LA-170, which stated that a "*study of solid sphere (Christy gadget) is being done by the magnetic method (Fowler) and X-ray (Fowler, Tuck . . .).*"

### D. *Critical Assembly Book*

Los Alamos historians gave a valuable perspective in their remarkable 1993 book, *Critical Assembly,*[16] but unfortunately, only in its classified version. We are making unclassified text from this source available to a broader readership. The book describes the patent and states that the invention was really Christy's. We quote from the classified version of *Critical Assembly*, pp. 363–4, and its footnote 173:

*"As Christy explained the discovery in the 'Record of Invention' for his 1946 patent application, 'Since even with these compromises, symmetry was still uncertain, it seemed desirable to examine theoretically the arrangement which was the limit of the thicker and thicker shells, and which seemed to be least subject to asymmetry, i.e. the solid gadget. It seemed clear that since in a solid arrangement the active material starts off as near the center as it can be placed, the motion of the active material is minimized and if it moves little it should be unable to develop asymmetries as large as the actual motion.' From estimates he made of the asymmetry, it appeared extremely likely that a solid gadget would not suffer from the major trouble of the hollow ones— namely asymmetry.*[173]*"*

"*Footnote-173: Updated Christy document in patent files, Solid Gadget, A-84-015, 4-16,17,18; Christy interview by Hoddeson, 4 April 1986, OH-117, T-Division progress Report, 31. Sept. 1944, LAMS-149, pp 2-4, 11. According to the patent application, Christy's conception took place prior to 2 Oct. 1944. The application includes a first sketch and a reference to notebook, A-147 . . . . The patent on the Christy Gadget was filed in Christy and Peierls' names on 16 Jan. 1946. Both Christy and Peierls confirm that, although Peierls' name is on the patent,* ***the idea was Christy's****; Peierls was likely included since he was the T-1 group leader and did calculations that showed feasibility. Peierls interviewed by Hoddeson, 20–21 March 1986, OH-111, and Christy interview by Hoddeson, 14 April 1986, OH-117.*"

We agree with the perspective that Peierls was included because he did useful calculations. But we doubt that his position as T-1 group leader played a role since the academic tradition at Los Alamos has always been such that authorship is earned through substantive technical contributions.

## 5. Open Sources

### A. *Hans Bethe Interview on YouTube*

Hans Bethe was the leader of the Theoretical Division during the Manhattan Project and was intimately familiar with the proposal and subsequent IBM calculations for the Christy Gadget. In a 1996 recorded set of interviews, available on YouTube,[25] Bethe said, "*Christy suggested . . . let the implosion act on a complete sphere of material; the implosion will compress it. . . . This Christy method was adopted and was tested in what is called Alamogordo in the New Mexican desert.*" As the immediate leader and long-term friend of Peierls,[26] Bethe's statement is compelling evidence for the role of Christy as the originator of the invention.

### B. *Edward Teller Memoirs*

Edward Teller's memoirs give sole credit for the idea of the Christy Gadget to Christy: [7]

"*Robert Christy pointed out that if we adopted a simpler plan, we would eliminate the difficulties* [asymmetries resulting from an implosion of a hollow sphere of plutonium]. *Instead of imploding a thin shell of plutonium, we could implode an assembly of the high explosives and tamper on a solid sphere of plutonium. With little movement of the plutonium, mixing would be practically impossible. The calculations showed that the converging power of the explosion, even in this conservative design, would produce enough compression for success. Footnote: That simple, ingenious suggestion made at an earlier time would have saved a great deal of effort. But the work was not wasted because it later helped in producing improved designs.*" (Ref. 7, p. 202)

This text, including Teller's footnote, makes it clear that the idea was not first proposed months earlier by someone else; a paragraph later, he refers to this novel idea as "*Christy's proposal.*" Furthermore, Teller usefully illustrates how the proposal generated new confidence in the implosion device. When James Conant, Harvard University and head of the National Defense Research Committee, was informed of Christy's new design, he muttered that "*this is the first time I have really thought it would work.*" (Ref. 7, p. 203)

The improved designs that Teller was referencing in his footnote were designs for levitated spheres and hollow shells. The hollow-boosting shell concept was successfully tested by Los Alamos in Operation Teapot (1955); Livermore subsequently made other useful design advances.

### *C. Robert Christy Interview on YouTube*

An interview with Christy is available on YouTube.[27] Again, Christy states his own role in inventing the gadget:

*"I suggested that instead of having a hollow shell which could implode in an irregular way, if you started with an essentially solid ball and then hit that with an explosive on the outside, that would be sure to stay like a sphere and therefore it could not fail. . . . It would be much more certain. This idea was bought as being the best way to proceed."*

Christy does not credit Peierls or anyone else as inventor.

### *D. Juliana Christy's Biography*

In Juliana Christy's hagiographic book about her late husband,[28] she quotes Robert Christy as stating:

*"The idea I had was to eliminate the hole in the middle of the implosion bomb and make it a solid sphere that would be compressed by the high explosives."*

### *E. Richard Rhodes' Book* **Dark Sun**

Richard Rhodes should not be expected to have unique insights into the questions explored in this paper. Nevertheless, his books on the history of the fission and fusion bomb developments are well regarded and thorough. Surprisingly, the details of the solid core invention are not given in *The Making of the Atomic Bomb*. Instead, they appear in *Dark Sun*. There, Rhodes credits only Robert Christy for proposing the use of a solid-plutonium ball.[29]

## 6. Conclusions

The exact details of the creative events in the Theoretical Division during the fall of 1944 will likely remain forever unknown. But Lorna Arnold's claim that the solid sphere design "*was Peierls' idea*" can be refuted. There is no evidence we are aware of that supports her claim (this is something of a surprise given her many impressive accomplishments as an historian). Quite the opposite: in an oral history interview at Los Alamos, Peierls himself said, "*Now, Christy's idea was to notice . . . That was a very novel thought. Now, how likely it is that someone else would have come up with that, I just don't know.*" (See Sec. 4B.) Peierls' writings never suggested otherwise.

Los Alamos' NSRC, along with numerous open source statements, provide convincing evidence that Christy invented the idea of the solid sphere implosion design for the Fat Man weapon. The new information that we have provided in Sec. 4 of this paper, presented as unclassified extracts from the classified originals, is (a) the patent evidence showing the evolution of the invention patent from 1945 (in Christy's name only) to 1946 (in Christy and Peierls' names); (b) our transcripts of the Los Alamos 1986 oral history interviews of Christy and Peierls, see Sec. 4B; (c) the Theoretical Division monthly progress report by Bethe and Peierls, Sec. 4C; and (d) the perspectives documented in the classified version of the 1993 book *Critical Assembly*, Sec. 4D.

In the classified version of *Critical Assembly* (unclassified extracts reproduced in Sec. 4D), Hoddeson et al. suggest that Peierls' name was added to the patent in 1946 because he conducted essential hydrodynamic calculations that showed it was feasible. We concur with this view from the Los Alamos historians. This scenario reflects a common phenomenon in science—that while the idea might have originated with one person, the process of working out the idea, refining it, performing the calculations, motivating key small-scale implosion experiments, and proving it to be feasible was a collaborative effort between Christy and Peierls. We speculate that after the war, Ralph Carlisle Smith was motivated to finalize and accurately clarify the many patents. As a fair-minded attorney, he wanted to attribute credit in the patent inventions as accurately as possible. Furthermore, we noted that his letter shown in Fig. 6 points to some input from government officials in Washington DC to add Peierls to the patent, for reasons that can only now be speculated on.

John von Neumann did, at an earlier time, suggest a design with a levitated solid ball, but it was in a different configuration than Christy's idea (von Neumann's had an air gap). The argument that Teller originated the Christy Gadget has also been refuted: Teller himself emphasized his own original insight was in the increased compression in implosion, but he credits the "ingenious idea" of a solid plutonium sphere to Christy (Sec. 5B).

Beyond the Christy gadget design, this paper has not addressed the origins of the implosion concept, except for mentioning the early interest by von Neumann, Teller, Tuck, and Neddermeyer. It is generally thought that in the U.S. the idea goes back to Richard Tolman as noted in Serber's *Primer*. Tom Kunkle is presently writing a paper on the German origins of implosion concepts, starting around 1940, and indeed whether any of this German work was known by US Manhattan Project researchers at the time as Lowell Wood has suggested[30].

The remarkable characters listed in our Dramatis Personae all played outsized roles in the development of the first atomic weapon. But as for the question of who was the originator of the Christy Gadget's invention, it has been a case of much ado about nothing. It was Christy.

## 7. Acknowledgment

We gratefully acknowledge useful discussions with Alan Carr, Madelaine Whitacre, Roger Meade, Lowell Wood, Frank Close, Sabine Lee, Nic Lewis, Bill Archer, Craig Carmer, and the AWE for their review, as well as very valuable input from Peter Christy (Bob's son). We also thank Dannie Alcazar for his assistance in locating documents in the NSRC, and Carl Gilbert for providing classification reviews. This is a Los Alamos unclassified document LA-UR-20-27638 version 3.

---

[1] A. B. CARR, "Thirty Minutes Before the Dawn: The Story of Trinity," *WRL* (this issue), *American Nuclear Society Nuclear Technology* (Submitted, 2021).

[2] E. N. BROWN, D. L. BOROVINA, "The Trinity High Explosives Implosion System: The Foundation for Precision Explosive Applications," *WRL* (this issue), *American Nuclear Society Nuclear Technology* (Submitted, 2021).

[3] A. B. CARR, "Documents Pertaining to the British Mission," *Los Alamos report* LA-UR-09-05504 (2009). R.C. Smith letter to Captain T.O. Jones, 18 September 1945.

[4] L. ARNOLD, *Britain and the H-Bomb*, Palgrave Macmillan, Basingstoke, Hampshire, (2001).

[5] F. CLOSE, *Trinity: The Treachery and Pursuit of the Most Dangerous Spy in History*, Allan Lane (2019).

[6] R. CHRISTY, Oral history interview with L. Hoddeson, April 14, 1986, OH-117, Los Alamos NSRC. Audio file Robert_Christy_2of4, at 12–15 minutes in.

[7] E. TELLER, *Memoirs: A Twentieth-Century Journey in Science and Politics*, Perseus Publishing, Cambridge, Massachusetts, 185 (2001).

[8] M. GOWING, *Britain and Atomic Energy 1939–1945*, Palgrave Macmillan, Basingatoke, London, 389-394 (1964) See MAUD in Appendix (p. 394) and Frisch-Peierls in Appendix (p. 389).

[9] J. R. OPPENHEIMER,"J.R. Oppenheimer letter to R. Peierls," Los Alamos document NSRC A-084-019 (1942).

[10] R. C. WILLIAMS and P. L. CANTELON, "Letters of J. Robert Oppenheimer," *The American Atom*, 36-37, University of Pennsylania Press, Philadelphia.

[11] B. ARCHER and N. MORGAN, ``On the origins of Lagrangian hydrodynamic methods'', this issue (ANS NT, submitted 2021)

[12] A. B. CARR, "Documents Pertaining to the British Mission," Los Alamos Report LA-UR-09-05504 (2009). These letters, and many others related to the British contributions, are available in a useful compilation by Alan Carr.

[13] J. LESTONE, this issue (ANS NT, submitted 2021); J. LESTONE and C. BATES, "The Bethe-Feynman Formula and WWII Atomic Bombs," Los Alamos report LA-20-00318 (2020).

[14] S. ULAM, "John von Neumann 1903–1957," *Bull. Amer. Math. Soc.*, 64, 1-49 (1958).

[15] R. MOORE, "Woolwich, Bruceton, Los Alamos: Monroe Jets and the Trinity Gadget," This issue. *American Nuclear Society Nuclear Technology* (Submitted, 2021).

[16] L. HODDESON, P. W. HENDRIKSEN, R. A. MEADE, and C. WESTFALL, *Critical Assembly, A Technical History of Los Alamos During the Oppenheimer Years 1943-1945*, Cambridge University Press (2008). The classified version has reference number CRM-HP-88-34 and is available on the ALDX Online Vault as BC01126607AN.

[17] R. C. SMITH, "British Mission," Los Alamos memorandum LAB-ADCS-127 (1949). Included in Ref. 12.

[18] F. M. SZASZ, *British Scientists and the Manhattan Project*, Macmillan Academic and Professional LDT (1992).

[19] A. WELLERSTEIN, Christy's Gadget: Reflections on a death," http://blog.nuclearsecrecy.com/2012/10/05/christys-gadget/ (2012).

[20] S. LEE, "'In no sense vital and actually not even important'? Reality and Perception of Britain's Contribution to the Development of Nuclear Weapons," *Contemporary British History*, 20, 2, 159-185 (2006), DOI: 10.1080/13619460600600680.

[21] R. CHRISTY and R. PEIERLS, "Method and Apparatus for Explosively Releasing Nuclear Energy," Los Alamos NSRC, 4-16, 17, 18, S-3956-X; A-84-015.

[22] T. A. CHADWICK and M.B. CHADWICK, ``Who Invented the Trinity Nuclear Test's Christy Gadget", Los Alamos unclassified report LA-UR-20-27638 version 2 (2020); Los Alamos WRL journal, Los Alamos report LA-CP-21-00045 (2021).

[23] R. PEIERLS, "Oral History Interview with L. Hoddeson," OH-111, audio tape 5 of 6, Los Alamos NSRC (March, 20-21, 1986).

[24] H. BETHE, "Theoretical Division Progress Report," LAMS-149. (September, 1944). After Bethe's introduction, each group leader provides a summary of their progress, including Peierls.

[25] H. BETHE, "Hans Bethe – Help from the British, and the 'Christy Gadget' (94/158)," *YouTube*, https://www.youtube.com/watch?v=i7KwPpXr1zU.

[26] S. LEE, *The Bethe-Peierls Correspondence,* World Scientific, Singapore, (2007).

[27] R. CHRISTY, "Robert Christy – Constructing the Nagasaki Atomic Bomb," *YouTube*, https://www.youtube.com/watch?v=Ez45QEMI5CA&list=PLVV0r6CmEsFyeOKI21OdxEh1TgQlnJ_Uj&index=8.

[28] I. J. CHRISTY, *Achieving the Rare: Robert Christy Journey in Physics and Beyond*, World Scientific, Singapore, (2013).

[29] R. RHODES, *Dark Sun: The Making of the Hydrogen Bomb*, Touchstone, New York, (1996).

[30] L. WOOD, private communication to MBC, 9/1/2020. Lowell Wood, a friend of Teller's, wrote: "*According to the story-as-told-by-Teller, the essence of the Idea arose in a 'casual' discussion one evening at his 'home' on Bathtub Row, soon after the Lab opened, when a just-arrived-in-town John von Neumann came over for dinner and the two of them fell to discussing fundamental alternatives to the 'Little Boy' concept, as a basic hedge against its then-already-well-understood-in-principle failure modes. Von Neumann tabled the spherical implosion concept, the outline of which he had heard from its seemingly-never-openly-published German origins around 1940 – which concept however was novel to Teller, who inquired for technical specifics, centered on then-attainable velocities and certain hydrodynamic complexities. Von Neumann then estimated-on-the-spot the quantitative aspects of interest to Teller, who apparently observed on-the-spot that then-remarkably high pressures would seemingly be realized at implosion culmination, which he commented were large compared to those at Earth's center (due to his earlier geophysics interests). The two of them then rather simultaneously realized that substantially-greater than zero-pressure solid-densities might thereby be realized in the vicinity of implosion*

*centers, which they proceeded to estimate – and to make seemingly-obvious inferences re. the 'impacts' on the critical masses of fast-fissile implosion-processed materials-of-interest. The following day, Teller related, he and von Neumann took the concept to Oppenheimer, who rather immediately directed the commencement of high-priority effort, for which effort Seth Neddermeyer was conscripted by Oppenheimer – and the story-from-there may be found in the set of LAMS documents bearing the names of one-or-more of Neddermeyer, Teller and von Neumann – as well as many others.*

*(Yes, I'm aware that this narrative is notably different from the long-since-Official Story – but it's the best that I have on-offer, and no one ``inside'' ever controverted its essentials in my hearing)*." (2020).

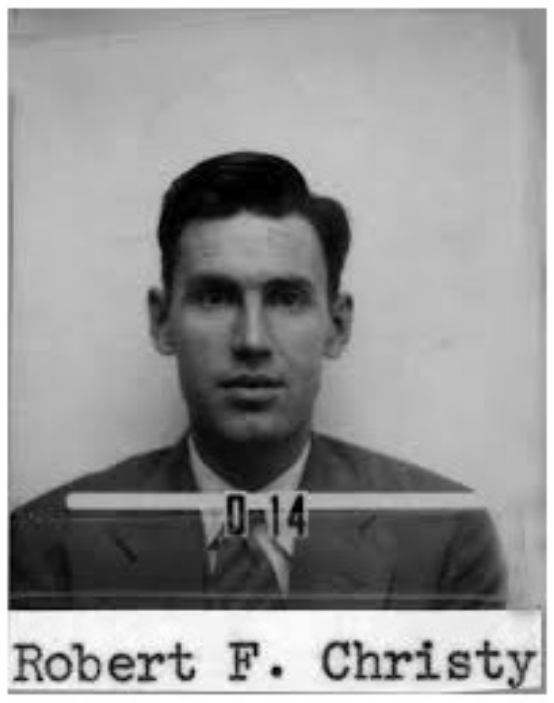

*Figure 1. Robert Christy photographed soon after his arrival in Los Alamos.*

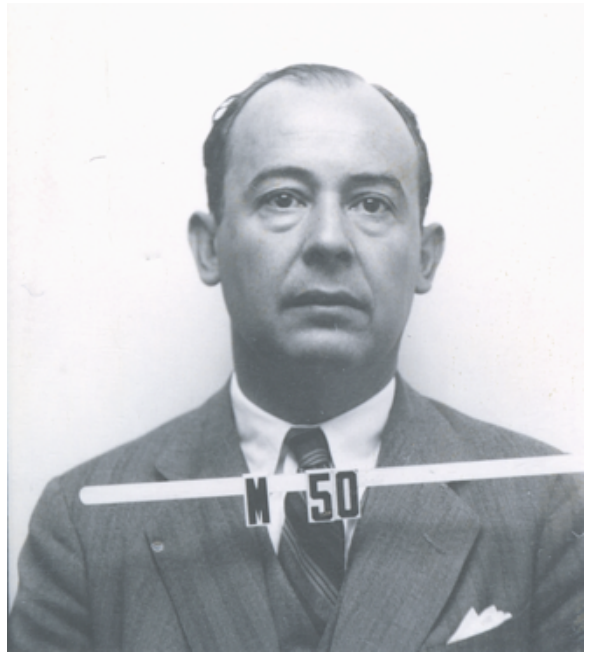

*Figure 3. John von Neumann.*

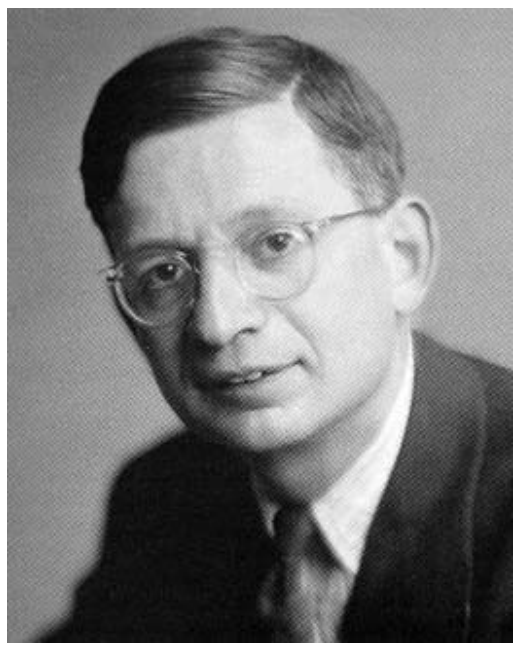

*Figure 2. Rudolf Peierls.*

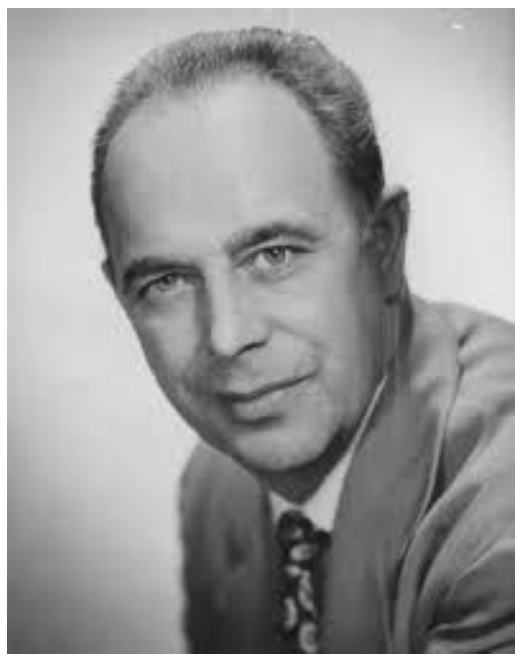

*Figure 4. Ralph Carlisle Smith.*

THEORETICAL DIVISION
MAY 10, 1945

H. A. Bethe, Division Leader, E-208, Ext 71
V. F. Weisskopf, Deputy Division Leader, E-217, Ext 74
J. von Neumann, Consultant, E-205½, Ext 72 R3

GROUP T-1
R. Peierls, Leader, E-119, Ext 178
R. F. Christy, Section Leader, E-120 Ext 178 R2
K. Fuchs, Section Leader, E-118 Ext 77
Baroody, E. M., E-121, Ext 178 R2
Calkin, J. W., E-117, Ext 77
Inglis, D. R., T-35, Ext 54
Keller, J., E-120, Ext 178 R2
Penny, W. G., E-101A, Ext 470
Podgor, T/5 S., E-116, Ext 469
Roberts, T/5 A. E., E-117, Ext 77
Skyrme, T.H.R., E-118, Ext 77
Stark, R. H., E-116, Ext 469
Stein, T/5 P. R., E-121, Ext 178

GROUP T-2
R. Serber, Leader, E-109, Ext 177
L. I. Schiff, Alt. Leader, E-108 Ext 76
Case, K. M., E-107, Ext 76
Glauber, R., E-107, Ext 76
Kurath, T/3 D., E-202, Ext 205
Rarita, W., E-111, Ext 177
Richman, C., E-110, Ext 177
Stehle, T/5 P., E-107, Ext 76

GROUP T-3
V. Weisskopf, Leader, E-217, Ext 74
R. E. Marshak, Alt. Leader, E-218 Ext 74 R2
Bellman, Pvt. R., E-222, Ext 468
Cohen, T/5 S., E-219, Ext 74 R2
Lennox, E., E-218, Ext 74 R2
Olum, Paul, E-216, Ext 74
Smith, J. H., E-219, Ext 74
Wing, Milton, E-222, Ext 468
Bowers, W. A., E-216, Ext 74

GROUP T-4
R. Feynman, Leader, E-206, Ext 72
J. Ashkin, Alt. Leader, E-209 Ext 72 R2
Ehrlich, R., E-210, Ext 72 R2
Peshkin, T/4 M., E-203, Ext 79 R3
Reines, F., E-210, Ext 72 R2
Welton, T. A., E-209, Ext 72 R2

GROUP T-5
D. Flanders, Leader, E-205 Ext 69 R1
P. Whitman, Alt. Leader, E-204 Ext 69 R2
Atkins, A. L., E-211, Ext 73 R2
Davis, R. R., E-215, Ext 74
de le Vin, E., E-201, Ext 79
Elliott, J., E-213, Ext 73 R1
Hauser, T/5 F. H., E-214, Ext 73 R1
Huber, T/4 D. C., E-120, Ext 70
Hudson, H., E-214, Ext 73 R1
Inglis, B., E-212, Ext 73 R2
Johnson, M., E-212 Ext 73 R2
Kellogg, T/5 H., E-203, Ext 69 R3
Langer, B., E-201, Ext 79
Page, T/3 W., E-214, Ext 73 R1
Rau, E. T/3., E-120, Ext 70
Staley, T/3 J., E-212, Ext 73 R2
Teller, M., E-214, Ext 73 R1
Vuletic, T/5 V., E-211 Ext 73 R2
Wilson, F., E-212, Ext 73 R2
Wright, T/5 E., E-213, Ext 73 R1
Young, T/Sgt., G., E-211, Ext 73 R2

GROUP T-6
E. Nelson, Leader, E-116,
N. Metropolis, Alt. Leader, E-115, Ext 78
R. Feynman, Consultant, E-206, Ext 72
Ewing, F. E., E-112, Ext 75
Goldberg, T/5., E-108, Ext 76
Kemeny, Pvt. J., E-105, Ext 75
Hamming, R. W., E-114, Ext 78
Heermans, Corp. A., E-105, Ext 75
Heller, T/5 A., E-105, Ext 75
Hurwitz, T/5 D., E-105, Ext 75
Johnston, T/3 J., E-103, Ext 75
Kington, T/4 J., E-105, Ext 75
Livesay, N., E-112, Ext 75
Ninger, H., E-105, Ext 75
Noah, F. E., E-105, Ext 75
Vorwald, T/5 A., E-105, Ext 75
Zimmerman, T/3 W., E-105, Ext 75

GROUP T-7
J. Hirschfelder, Leader, T-30 Ext 206
J. Magee, Alt. Leader, T-30, Ext 206
Brummer, T/5 E., T-28, Ext 206
Feckete, T/4 P., T-28, Ext 206
Larson, T/4 L., T-26, Ext 206
Ostrow, E., T-30, Ext 206
Schwartz, T/4 P., T-28, Ext 206

GROUP T-8
G. Placzek, Leader, E-220, Ext 468
Mark, C., E-221, Ext 468 R2
Carlson, B., E-221, Ext 468 R2
Day

*Figure 5. An image of the Theoreotical Division Organizational Chart during the Manhattan Project, May 10 1945. There were 18 researchers at Los Alamos during the Manhattan Project who either had, or would be later given, Nobel Prizes. T-Division had four of them (Bethe, Glauber, Reines and Feynman).*

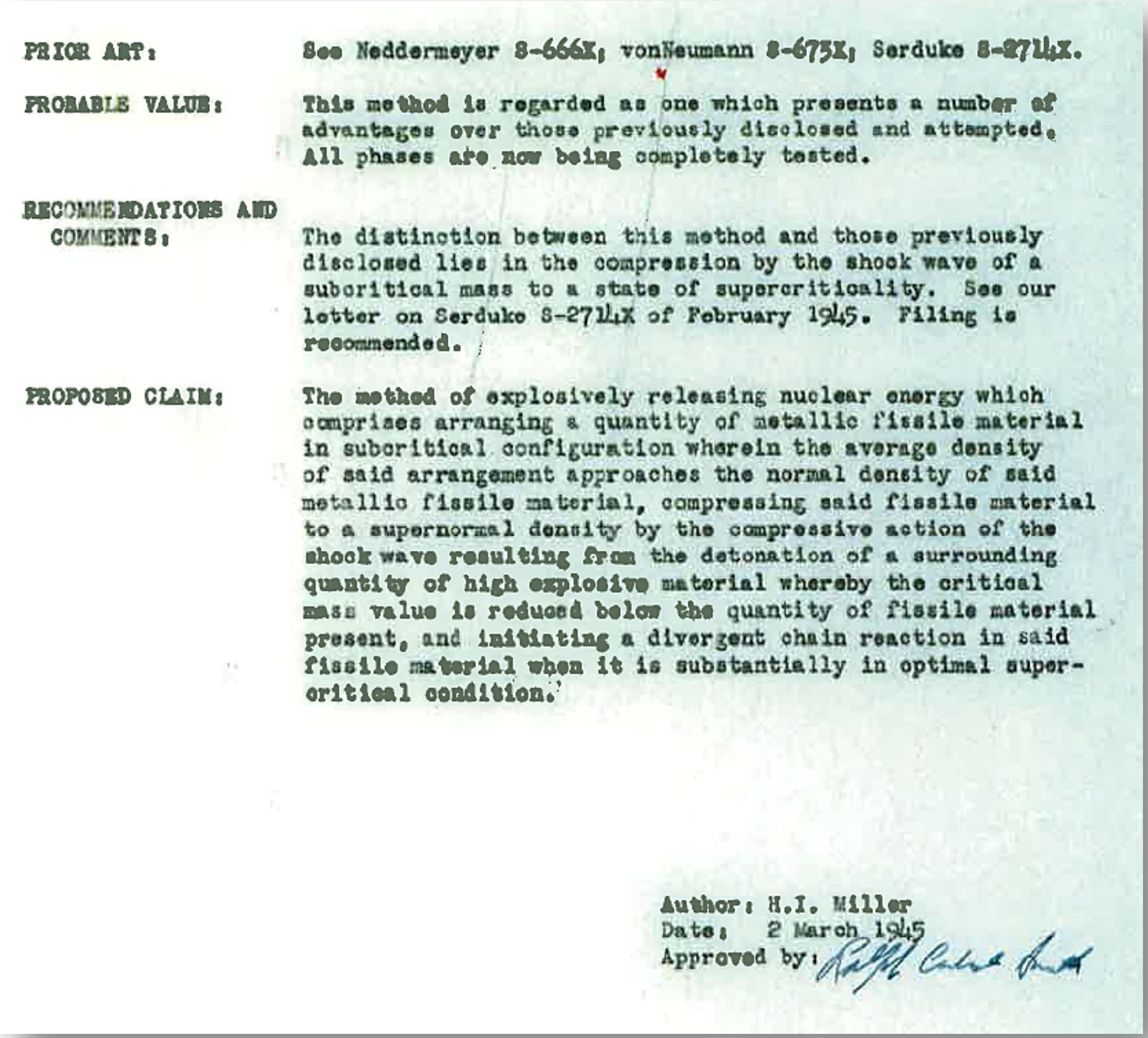

PRIOR ART: See Neddermeyer S-666X; vonNeumann S-673X; Serduke S-2714X.

PROBABLE VALUE: This method is regarded as one which presents a number of advantages over those previously disclosed and attempted. All phases are now being completely tested.

RECOMMENDATIONS AND COMMENTS: The distinction between this method and those previously disclosed lies in the compression by the shock wave of a subcritical mass to a state of supercriticality. See our letter on Serduke S-2714X of February 1945. Filing is recommended.

PROPOSED CLAIM: The method of explosively releasing nuclear energy which comprises arranging a quantity of metallic fissile material in subcritical configuration wherein the average density of said arrangement approaches the normal density of said metallic fissile material, compressing said fissile material to a supernormal density by the compressive action of the shock wave resulting from the detonation of a surrounding quantity of high explosive material whereby the critical mass value is reduced below the quantity of fissile material present, and initiating a divergent chain reaction in said fissile material when it is substantially in optimal supercritical condition.

Author: H.I. Miller
Date: 2 March 1945
Approved by:

*Figure 6. An image of another extract from the original March 6, 1945, patent with Robert Christy as the sole author. This extract references other related, but different, inventions patents by Neddermeyer, von Neumann, and Serduke that are also held in Los Alamos' NSRC, see Sec. 3. It also details the Proposed Claim. In Fig. 7 we show an example of the claim in the final Christy-Peierls patent, that is very similar.*

It is claimed:

28. The process of explosively releasing nuclear energy which comprises arranging a quantity of normal density fissile material in sub-critical solid spherical configuration possessing a low neutron background within and in contact with a layer of tamper material, positioning a quantity of high explosive material substantially surrounding said tamper material, said explosive material having no substantial volume of free space within its confines, detonating said high explosive material whereby said tamper and fissile material are subjected to an implosive shock and are compressed to a supernormal density and supercritical condition at a rate at which there is little probability of neutron initiation of the chain reaction before the attainment of optimal supercriticality, and initiating a neutron chain reaction by introducing neutrons into the fissile material when it is in the optimal supercritical condition.

29. Each and every novel feature and combination of novel features present in or possessed by the method or apparatus herein disclosed.

*Figure 7. This image is from the (now joint) Christy-Peierls patent from January 1946, showing items 28 and 29, which follow "It is claimed:" There are 28 such paragraphs, all involving just small variations of the same claim. An earlier similar version of the claim is shown in Fig. 6, for the Christy-only patent.*

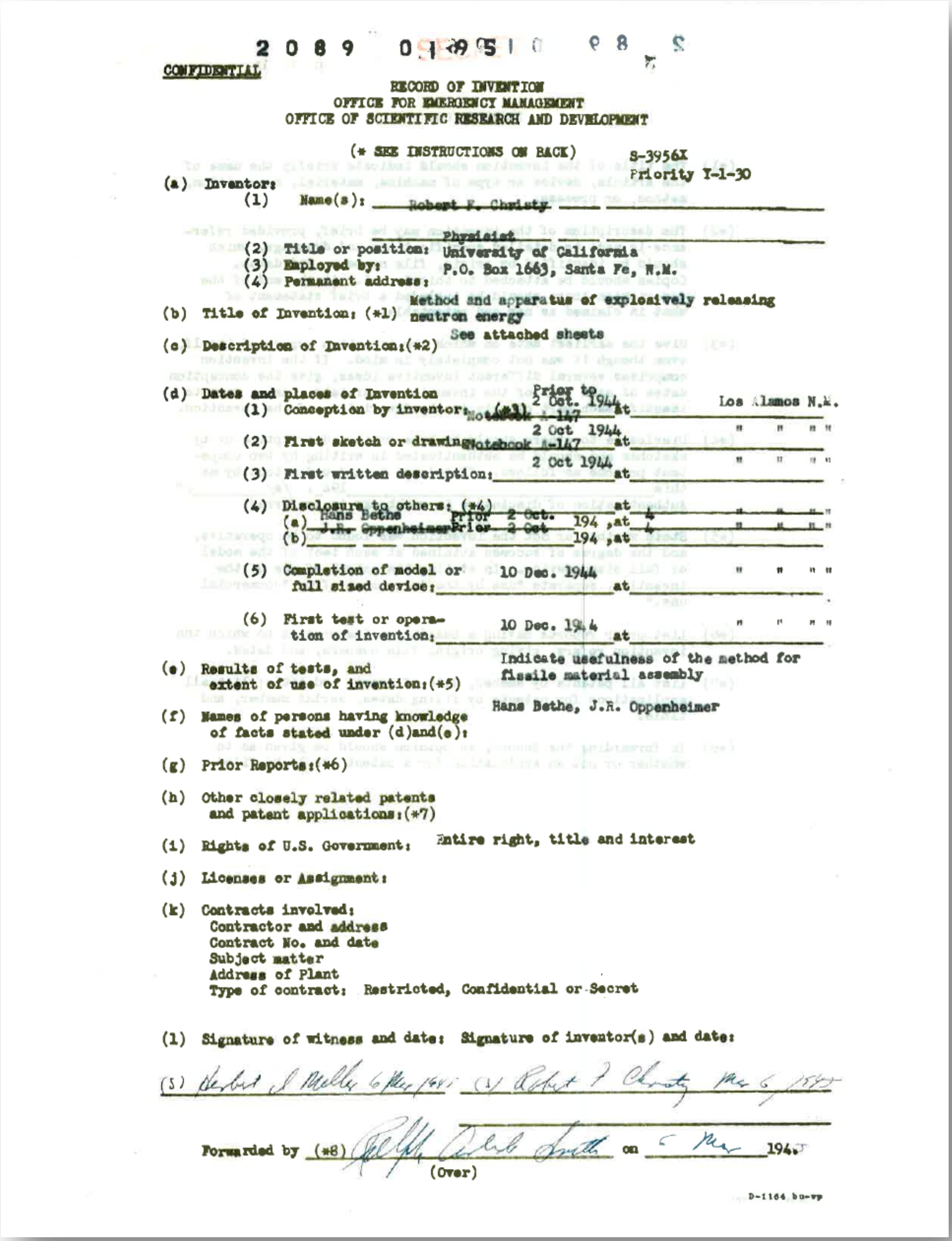

2089 01951 0 98 2

CONFIDENTIAL

RECORD OF INVENTION
OFFICE FOR EMERGENCY MANAGEMENT
OFFICE OF SCIENTIFIC RESEARCH AND DEVELOPMENT

(* SEE INSTRUCTIONS ON BACK)

S-3956X
Priority Y-1-30

(a) Inventor:
(1) Name(s): Robert F. Christy
(2) Title or position: Physicist
(3) Employed by: University of California
(4) Permanent address: P.O. Box 1663, Santa Fe, N.M.

(b) Title of Invention: (*1) Method and apparatus of explosively releasing neutron energy

(c) Description of Invention:(*2) See attached sheets

(d) Dates and places of Invention
(1) Conception by inventor: (*3) Prior to 2 Oct. 1944 Notebook A-147 at Los Alamos N.M.
(2) First sketch or drawing: 2 Oct 1944 Notebook A-147 at " " " "
(3) First written description: 2 Oct 1944 at " " " "
(4) Disclosure to others: (*4) at
(a) Hans Bethe Prior 2 Oct. 194 ,at " " " "
(b) J.R. Oppenheimer Prior 2 Oct 194 ,at " " " "
(5) Completion of model or full sized device: 10 Dec. 1944 at " " " "
(6) First test or operation of invention: 10 Dec. 1944 at " " " "

(e) Results of tests, and extent of use of invention:(*5) Indicate usefulness of the method for fissile material assembly

(f) Names of persons having knowledge of facts stated under (d)and(e): Hans Bethe, J.R. Oppenheimer

(g) Prior Reports:(*6)

(h) Other closely related patents and patent applications:(*7)

(i) Rights of U.S. Government: Entire right, title and interest

(j) Licenses or Assignment:

(k) Contracts involved:
Contractor and address
Contract No. and date
Subject matter
Address of Plant
Type of contract: Restricted, Confidential or Secret

(l) Signature of witness and date: Signature of inventor(s) and date:

(S) Herbert I Miller 6 Mar 1945 (S) Robert F. Christy Mar 6, 1945

Forwarded by (*8) Ralph Carlisle Smith on 6 Mar 1945

(Over)

D-1164 bu-vp

*Figure 8. An image of an extract from the original March 6, 1945, patent with Robert Christy as the sole author/inventor, showing the Record of Invention. It is signed by Ralph Carlisle Smith, the person in charge of patent applications at Los Alamos. Note that it refers to the conception as "prior to 2 Oct. 1944" and references the first sketch or drawing in Christy's laboratory "Notebook A-147." Bethe's Theoretical Division monthly progress report LAMS-149 from September 1944 also references Christy's idea (see Fig. 11). Note the typo in the title ("neutron energy" instead of "nuclear energy," not present in the original handwritten version in Fig. 9.*

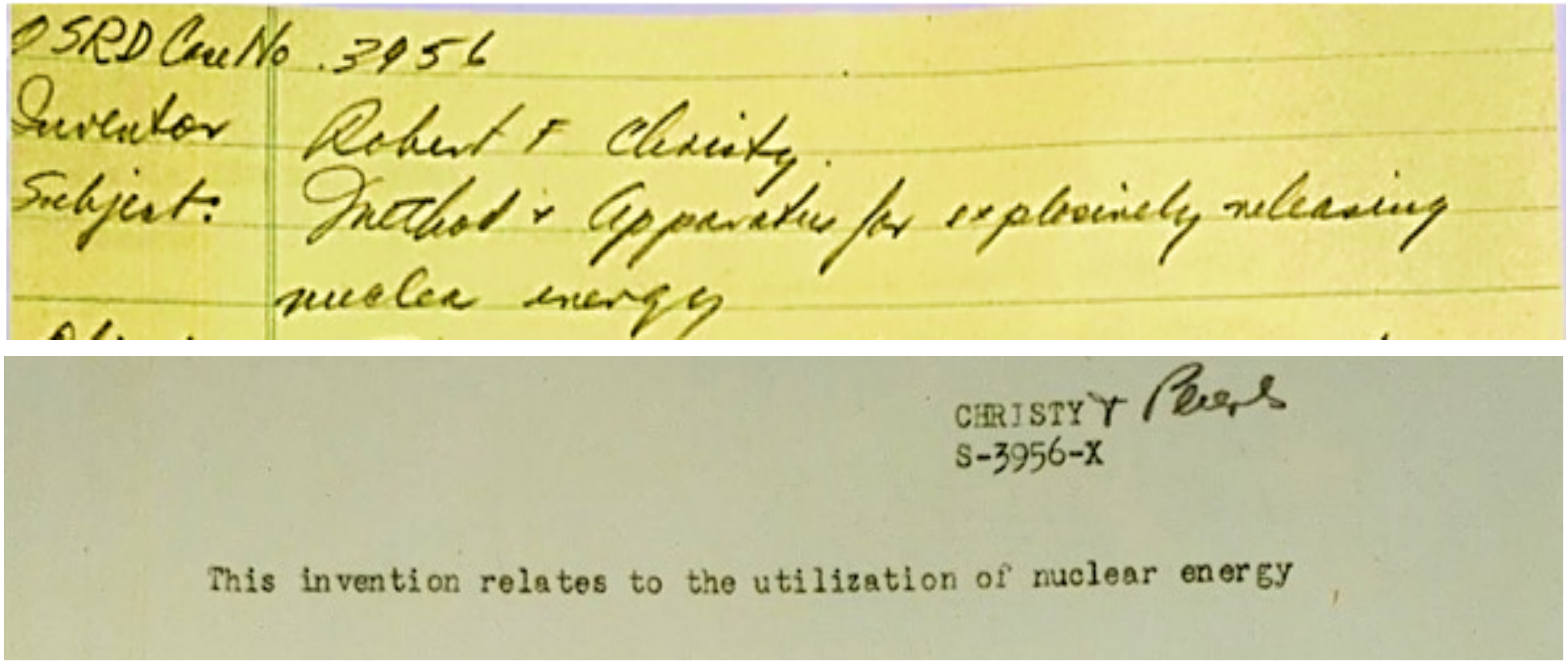
OSRD Case No. 3956
Inventor Robert F Christy
Subject: Method & Apparatus for explosively releasing nuclear energy

CHRISTY & Peierls
S-3956-X

This invention relates to the utilization of nuclear energy

*Figure 9. The upper image is from the early 1945 Christy-only authored patent. The lower image comes from the point at which Peierls was added as a co-inventor (Peierls added his name as an author, by hand).*

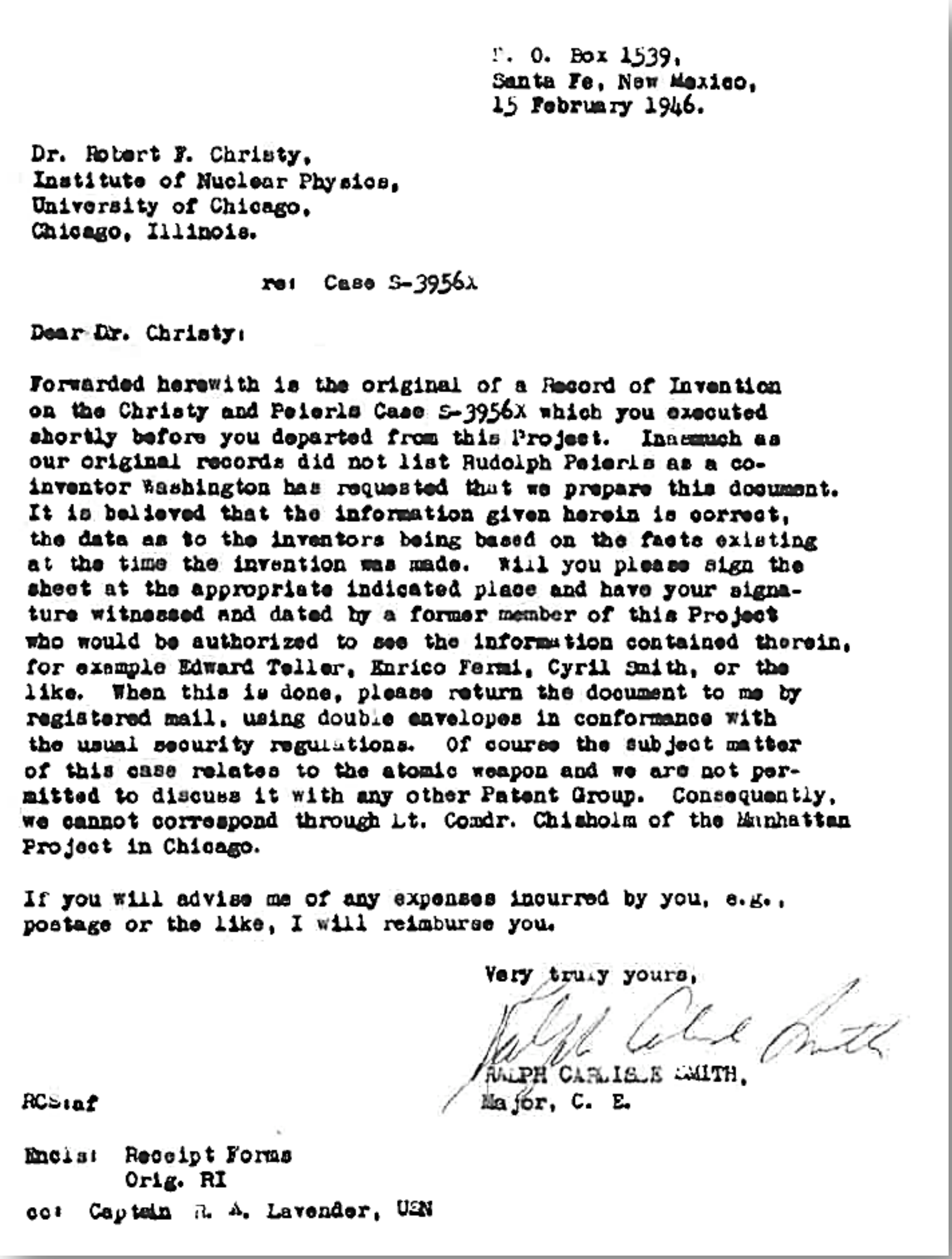

P. O. Box 1539,
Santa Fe, New Mexico,
15 February 1946.

Dr. Robert F. Christy,
Institute of Nuclear Physics,
University of Chicago,
Chicago, Illinois.

re: Case S-3956X

Dear Dr. Christy:

Forwarded herewith is the original of a Record of Invention on the Christy and Peierls Case S-3956X which you executed shortly before you departed from this Project. Inasmuch as our original records did not list Rudolph Peierls as a co-inventor Washington has requested that we prepare this document. It is believed that the information given herein is correct, the data as to the inventors being based on the facts existing at the time the invention was made. Will you please sign the sheet at the appropriate indicated place and have your signature witnessed and dated by a former member of this Project who would be authorized to see the information contained therein, for example Edward Teller, Enrico Fermi, Cyril Smith, or the like. When this is done, please return the document to me by registered mail, using double envelopes in conformance with the usual security regulations. Of course the subject matter of this case relates to the atomic weapon and we are not permitted to discuss it with any other Patent Group. Consequently, we cannot correspond through Lt. Comdr. Chisholm of the Manhattan Project in Chicago.

If you will advise me of any expenses incurred by you, e.g., postage or the like, I will reimburse you.

Very truly yours,

Ralph Carlisle Smith

RALPH CARLISLE SMITH,
Major, C. E.

RCS:af

Encls: Receipt Forms
Orig. RI

cc: Captain R. A. Lavender, USN

*Figure 10. This image shows Ralph Carlisle Smith's February 1946 letter to Christy, where a decision has been made to add Peierls' name as co-inventor to the patent.*

Extract LAMS-149.. Bethe intro:

As a possible further insurance against failure to achieve symmetry in an implosion of more standard design, we have also considered the possibility of imploding an initially nearly solid sphere. Some compression is expected even in this case

Extract Peirls

GROUP T-I, R. E. PEIERLS, GROUP LEADER

Implosion of a Solid Sphere

Christy has recently made some estimates on the performance of a solid implosion gadget which operates on the compression achieved by a converging shock wave in solid material. The object of this calculation was twofold; first to find what efficiency could be achieved by an arrangement which probably avoids the difficult problem of obtaining high symmetry and could probably be built with our present knowledge. The result of

*Figure 11. Images from the September 1944 Theoretical Division Progress Report by Hans Bethe, document LAMS-149. These remarkably informative monthly progress reports follow a standard format: Bethe provides an introduction, and then each group leader provides a summary of their group's work. These extracts show the first Los Alamos report documentation of Christy's invention of his gadget idea, in Bethe's and Peierls' words.*